\documentclass[aps,prl,preprint,groupedaddress]{revtex4-1}

\usepackage{lineno}

\usepackage{graphicx}
\usepackage{epstopdf}
\usepackage[utf8]{inputenc}
\usepackage{amsmath}
\usepackage[T1]{fontenc}
\usepackage{lmodern}
\begin{document}


\title{Closing the superconducting gap in small Pb-nanoislands with high magnetic fields}


\author{Steffen Rolf-Pissarczyk}
\email[]{steffen.rolf-pissarczyk@mpsd.mpg.de}
\author{Jacob A.J. Burgess}
\author{Shichao Yan}
\author{Sebastian Loth}
\email[]{sebastian.loth@mpsd.mpg.de}
\affiliation{Max Planck Institute for the Structure and Dynamics of Matter, Luruper Chaussee 149, 22761 Hamburg, Germany }
\affiliation{Max Planck Institute for Solid State Research, Heisenbergstraße 1, 70569 Stuttgart, Germany}



\date{\today}

\begin{abstract}
Superconducting properties change in confined geometries. Here we study the effects of strong confinement in nanosized Pb-islands on Si(111) 7$\times$7. Small hexagonal islands with diameters less than 50 nm and a uniform height of 7 atomic layers are formed by depositing Pb at low temperature and annealing at 300 K. We measure the tunneling spectra of individual Pb-nanoislands using a low-temperature scanning tunneling microscope operated at 0.6 K, and follow the narrowing of the superconducting gap as a function of magnetic field. We find the critical magnetic field, at which the superconducting gap vanishes, reaches several Tesla, which represents a greater than 50-fold enhancement compared to the bulk value. By independently measuring the size of the superconducting gap, and the critical magnetic field that quenches superconductivity for a range of nanoislands we can correlate these two fundamental parameters and estimate the maximal achievable critical field for 7 ML Pb-nanoislands to 7 T.
\end{abstract}


\maketitle

\section{INTRODUCTION}
Nanostructures in which superconductivity and magnetism can coexist give rise to interesting phenomena such as Yu-Shiba-Rusinov states within the superconducting gap,\cite{Ji2010,Yazdani1997,Hudson2001} competition with Kondo screening \cite{Franke2011} and Majorana bound states.\cite{Nadj-Perge2014,Peng2015} Tuning the properties of nanoscale magnetic systems commonly requires high magnetic fields, however those same fields generally quench superconductivity at small amplitudes. Therefore there is a need for superconducting substrates that can be engineered to have high enough critical magnetic fields to permit convenient manipulation of the magnetic atoms or nanostructures. A promising avenue is nanoscale spatial confinement, which can  drastically alter the properties of conventional Bardeen-Cooper-Schrieffer, BCS, superconductors and can enhance the critical fields by several orders of  magnitude.\cite{Eltschka2015,Li2003}
\par
Here we focus on the impact of strong three-dimensional spatial confinement in Pb nano-islands on Si(111) 7$\times$7. Quantum size effects in this system influence growth and morphology and – depending on the growth parameters – lead to extremely flat thin films or well-defined islands.\cite{Chang2002,Su2001,Khajetoorians2009} Pb films have been found to be superconducting starting at one monolayer thickness with a critical magnetic field of 0.2 T, approximately a factor two higher than that of bulk Pb.\cite{Brun2014,Qin2009} But thin Pb films show vortex formation leading to non-uniform magnetic fields at the superconductor surface.\cite{Cody1968,Dolan1973,Ning2010} In Pb islands vortex formation becomes confined and eventually suppressed completely.\cite{Cren2009,Schweigert1998,Nishio2008} In this regime the critical field perpendicular to those islands starts to increase due to less effective screening of the magnetic field.\cite{Li2003,Schmidt1997} Critical magnetic fields in excess of 1 T have been observed in conjunction with a gradual suppression of superconductivity that varies with island size and thickness.\cite{Cren2009,Nishio2008,Eom2006,Brun2009,Kim2011} By further reducing the superconducting island size their electronic structure becomes discrete and shell-effects, electron number or quantum well effects become significant.\cite{Li2003,Bose2009,Bose2014} Earlier studies of superconducting particles in this size regime focused on the interaction of multigrain systems where nearby superconducting particles influenced each other and proximity effects play an important role.\cite{Sternfeld2005,Abeles1967}
\par
To investigate the limit of tuning superconductivity by means of geometrical confinement we prepared very small isolated Pb-islands on Si(111) 7$\times$7 with sizes of less than 50 nm in any dimension, smaller than those studied previously. The smallest islands investigated are close to the size at which superconductivity breaks down (see supplemental information).\cite{Anderson1959} The islands are well-isolated, oblate objects with a thickness of 7 atomic layers (ML), and an approximately hexagonal shape, Figure 1. The volume, V, of the islands was between 160 nm$^3$ and 4700 nm$^3$.
\par
Low-temperature scanning tunneling microscopy is used to address the properties of individual superconducting islands. We measure the dependence of the superconducting tunneling gap on island size and magnetic field by recording tunneling spectra of the differential conductance. Quenching of the superconducting state is directly observed in these spectra by closing of the superconducting gap when the critical field is reached. Magnetic fields up to 9 T were applied perpendicular to the sample surface and thereby access a magnetic field regime approximately five times higher than reported previously.\cite{Nishio2008} With this method we can experimentally observe that very small islands with diameter of 10 nm support superconductivity up to $(5.2\pm 0.4)$ T. By correlating zero-field gap and critical magnetic field for all islands studied in this experiment, we can estimate the maximum achievable critical field to be $6.7_{-1.5}^{+1.8}$ T.
\section{RESULTS AND DISCUSSION}
\begin{figure}
\includegraphics{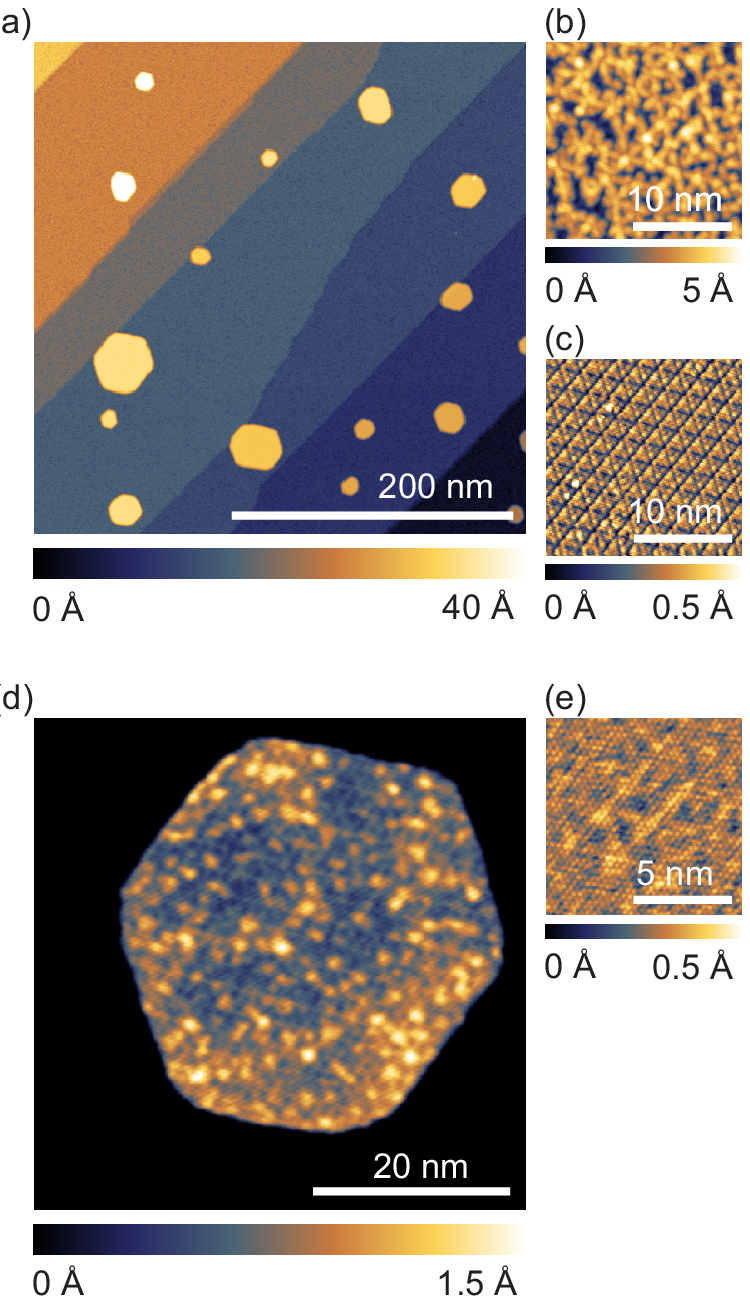}%
\caption{\label{fig1} Constant current STM topographies of nanosized Pb-islands. (a) The Si(111) 7$\times$7 surface is covered with a Pb wetting layer on which the hexagonal Pb-islands are grown ($V = 0.1$ V, $I = 70$ pA). The visible atomic steps and flat terraces result from the underlying Si surface. (b) Zoom-in to wetting layer (V = 0.01 V, I = 150 pA). The wetting layer exhibits a some residual structure of the Si(111) 7$\times$7 substrate, but otherwise appears disordered. (c) Constant current topography of the Si(111) 7$\times$7 surface prior to Pb evaporation ($V = -2.2$ V, $I = 100$ pA). (d) The top surface of a typical Pb-island. The impact of the mostly disordered wetting layer is observable ($V = 0.1$ V, $I = 2.5$ nA). (e) Zoom-in showing the hexagonal crystalline structure of the top surface of a Pb-island ($V = -1.2$ mV, $I = 1$ nA).}
\end{figure}
Small flat well-separated Pb islands are formed with low-temperature deposition of Pb onto a cold Si(111) 7$\times$7 reconstructed surface at approximately 100 K, and are subjected to subsequent annealing at room temperature for 10 min, Figure \ref{fig1} (for details see Methods). The Pb islands are located on a Pb wetting layer, which is also superconducting at 0.6 K (see SI). The wetting layer exhibits some residual structure  with residual order reminiscent of the 7$\times$7 reconstruction, Figure \ref{fig1}b-c. We attribute the disordered appearance of the wetting layer to the cold sample deposition which reduces diffusion and suppresses the formation of an ordered wetting layer that was found for Pb deposition at higher temperatures.\cite{Hupalo2007,Seehofer1995,Chan2003} The island height varies between six and eight atomic layers, however we restrict our measurements to islands with 7 ML height which was the most common height. In contrast to the disordered wetting layer the islands themselves have a hexagonally ordered surface indicating that they are crystalline Pb (Figure 1d). The shape of the island allows us to express the island volume, $V$, as an effective diameter, $d=2\sqrt{V/(7h_{ML} \pi)}$ with $h_{ML} = 0.285$ nm, the height of one atomic layer of Pb.\cite{Budde2000}. The disorder between the island and Si(111) substrate plays an important role for the superconducting properties of the islands. It limits the quasiparticle mean free path and brings it into the dirty or diffusive limit.\cite{Cren2009} The defect density was found to be the same for all measured islands. Hence variations of superconducting properties observed between islands of different size cannot be attributed to disorder but must stem from size effects. 

\par
The superconducting gaps of different Pb-islands are measured with scanning tunneling spectroscopy (for details see Methods). The differential conductance spectra, $dI/dV$, show a superconducting gap centered at 0 V, Figure 2a. No pronounced coherence peaks are observed for the small islands, whereas they appear for larger islands (see Figure S1).
\begin{figure}
\includegraphics{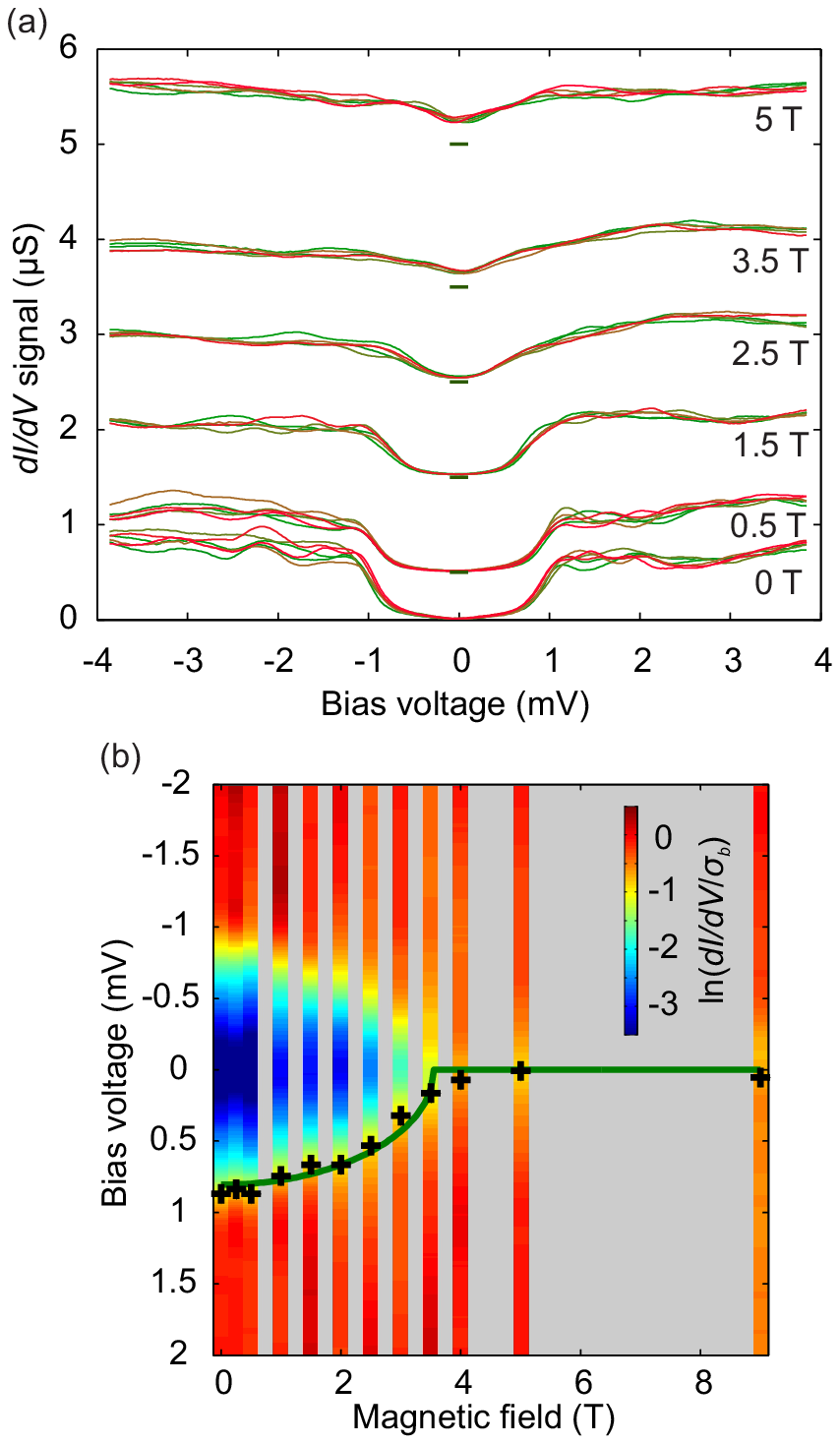}%
\caption{\label{fig2} (a) The superconducting gap in spectra of the differential conductance ($dI/dV$) taken at different magnetic fields on a Pb-island with diameter $d = 22$ nm as function of bias voltage. Spectra acquired at different magnetic fields are shifted vertically for clarity. Each bar indicates the zero conductance at the corresponding magnetic field. The $dI/dV$ spectra are averaged over different positions on the island which is indicated by different colors. At zero field a fit according to the Dynes formula is performed (blue dotted line), $\varDelta_0 = (1.03 \pm 0.01)$ meV.  In the further analysis the $dI/dV$ spectra of different positions are averaged and normalized to the averaged background conductance, $\sigma_b$. (b) The averaged  $dI/dV$ spectra show the closing of the superconducting gap and are plotted color-coded as function of bias voltage and magnetic field, and with a logarithmic color scale. The grey area indicates magnetic fields at which no data was recorded. The black crosses mark the position of the bias voltage width $V_e$ at which the normalized conductance dropped to $1/e$ of the background conductance. The evolution of the superconducting gap (green line) is fitted to $V_e$ using equation (\ref{deltaB}). For this island, a critical field of $B_c = (3.5 \pm 0.1)$ T was found.}
\end{figure}
\par
We find that conductance spectra recorded at different positions on an island vary slightly for voltages exceeding the superconducting gap. These fluctuations likely have their origin in the disordered wetting layer that is between the island and the Si surface. This is corroborated by the variations in the apparent topographic height on the Pb-island surface which shows a similar structure to that of the wetting layer, Figure 1b. To remove such localized influences from the spectrum we average spectra taken at four to six different positions to obtain a representative spectrum of each island.
\par
Applying a magnetic field perpendicular to the surface the $dI/dV$ spectra show that the gap closes with increasing magnetic field. The gap vanishes at approximately 3.5 T for the island shown in Figure 2a. The closing of the gap is similar for the different positions measured on the island and no sign of vortex formation was found on this island or any other studied here, all islands had an effective diameter less than 54 nm. This matches to previously reported suppression of vortex formation in Pb islands below 60 nm diameter \cite{Nishio2008} and indicates that the superconducting state is quenched uniformly at one critical field for each island without prior vortex formation.  The conductance at zero bias exhibits residual variation above the critical field which we don’t associate with superconductivity of the island  because it persists on all islands up to the highest magnetic field we could apply (9 T). Besides other possibilities such as a small variation in the differential conductivity of the tip, this tail could originate from fluctuation effects.\cite{Brihuega2011}
\par
By systematically recording $dI/dV$ spectra between 0 T and 9 T we can directly observe at which magnetic field the superconducting gap vanishes. To determine a quantitative value for the critical field, $B_c$, we extract the width of the superconducting gap of each spectrum and fit its evolution with magnetic field with the functional form expected for the superconducting order parameter. In the framework of the Ginzburg-Landau Theory, and for sufficiently small superconductors, the order parameter in an externally applied magnetic field is given by:\cite{Schmidt1997}

\begin{equation}
\varDelta(B)=\varDelta_0 \sqrt{ 1- \Big( \frac{B}{B_c} \Big)^2 } \label{deltaB}
\end{equation}

Where $\varDelta_0$ is the zero-field gap, $B$ is the magnetic field, and $B_c$ is the critical magnetic field at which superconductivity vanishes.
\par
At zero magnetic field the $dI/dV$ spectra fit well to the shape expected of the tunneling spectrum for a BCS superconductor and we can extract $\varDelta_0$ by fitting the spectrum with the Dynes formula \cite{Dynes1978} (see Fig. S1). We find that extraction of $\varDelta(B)$ from fitting the $dI/dV$ spectra is not feasible in large magnetic fields.  Although we have no indication that the spectra should deviate from a BCS-like behavior, analytical descriptions of superconducting tunneling spectra such as the Dynes equation or Maki’s theory \cite{Eltschka2015} which work well  at zero magnetic field, fail to reproduce the measured spectra at high magnetic field. Therefore we chose purely empirical method to extract the evolution of the gap width with magnetic field based on the experimental data. First, we determined the average tunnel junction conductance, $\sigma_b$, for each magnetic field as the average of the conductance for voltages significantly larger than the gap, $|V| > 2$ mV. Then, we define an effective gap width, $V_e$, where $V_e$ is the voltage at which the differential conductance exceeds $\sigma_b/e$. We chose this threshold value because the gap closing is clearly observable if the normalized averaged conductance spectra are plotted with a logarithmic scale (Fig 2). We note that it is sufficient to measure the magnetic field dependence of an observable that is proportional to the superconducting gap in order to determine the critical magnetic field. We verified numerically that $V_e$ is, in good approximation, proportional to the superconducting gap size, (see SI). Hence, substituting with $V_e$ in equation (\ref{deltaB}) and fitting it to the experimentally determined $V_e(B)$ yields the same critical field $B_c$ as fitting of $\varDelta(B)$ would yield, but now the fitting process is independent of choice of a model for the superconducting tunneling spectrum. We find that the observed evolution of $V_e$ with magnetic field fits well to the expected behavior for the order parameter, i.e. the superconducting gap $\varDelta(B)$. For the Pb-nanoisland shown in Figure \ref{fig2}b, $V_e$ is 0.9 mV at 0~T, changes only gradually up to 2~T and then reduces quickly to 0 mV when approaching 3.5~T
\par
\begin{figure*}
\includegraphics{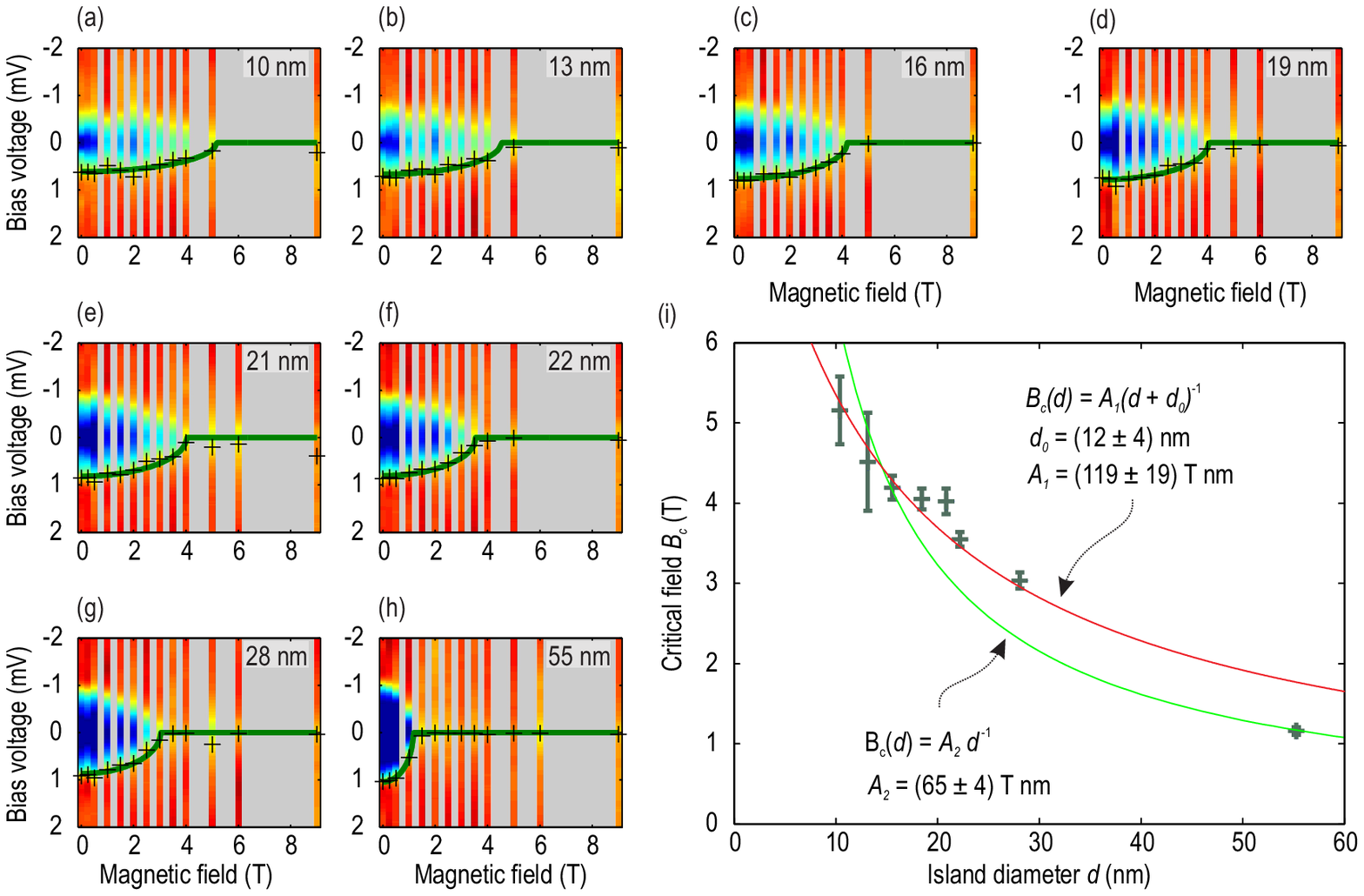}%
\caption{\label{fig3} (a-h) Color-coded $dI/dV$ spectra for islands with different diameters d from 55 nm to 10 nm as a function of bias voltage and magnetic field. The data was treated and plotted in the same fashion as demonstrated in Figure \ref{fig2}b. (i) The fitted critical fields, $B_c$, as function of the diameter d of the islands showing that $B_c$ increases as $d$ decreases. The highest observed critical field is $(5.2 \pm 0.4)$ T. The trend of the critical field is fitted by two different functions $A_2 /d $ (green line) and $A_1 /(d_0+d)$ (red line).}
\end{figure*}
To determine the effect of island size on $B_c$, we performed the same measurement and analysis on seven other Pb-islands. The island size varied from 55 nm to 10 nm in diameter. No island exhibited vortices in agreement with the results of Nishio et al.\cite{Nishio2008} The series of islands showed an increase of critical field from $(1.16 \pm 0.05)$ T for the 55 nm sized island up to $(5.2 \pm 0.4)$ T for the 10 nm island, Figure \ref{fig3}. 
\par
The resulting dependence of the critical field with island diameter is shown in Figure \ref{fig3}i. Previous reports found a $1/d$ trend for such superconducting particles in agreement with Ginzburg-Landau theory.\cite{Li2003,Schmidt1997} Whereas we observe the general trend of increasing critical fields with decreasing diameter, the fit to a $1/d$ trend shows deviations which are larger than the errors of our measurement.  This may reflect an effective  size of the superconducting entity that does not necessarily match to the geometrical size of the island because of coupling between island and wetting layer. A significantly better fit to the measured critical field is obtained by fitting with $1/(d+d_0)$.  Here we assume that the contribution $d_0$ is independent of island size and magnetic field. The fit diameter offset $d_0$ has a value of $(+12 \pm 4)$ nm and increases the effective superconducting area of each island compared to their geometric area. We therefore conclude that the superconducting proximity effect in the wetting layer contributes significantly and has a larger influence than inverse proximity effect in the island.\cite{Cren2009}
\par
In the absence of magnetic fields such proximity-induced superconductivity in a metallic or in a superconducting layer has been reported for Pb-islands on Si(111) with a decay length similar to the value of $d_0$ found here.\cite{Serrier-Garcia2013,Kim2012} Furthermore, experiments performed with Pb-islands surrounded by a metallic wetting layer showed that the superconducting proximity effect also perseveres in magnetic fields.\cite{Caminale2014} Our study suggests that the induced superconductivity in the wetting layer persists at magnetic fields that significantly exceed the critical field of the wetting layer, which is approx. 0.2 T.\cite{Brun2014} This is reasonable since earlier studies observed that such induced superconductivity in the wetting layer also persists above the transition temperature of a surrounding superconducting layer.\cite{Cherkez2014}
It is worth noting that the assumption of a constant decay length of the proximity effect for different islands and magnetic fields is a simplification. A more sophisticated theoretical model would be needed to account for the different contributions like the inverse proximity effect as well as altered coupling parameters for different island geometries. 
\par
\begin{figure}
\includegraphics{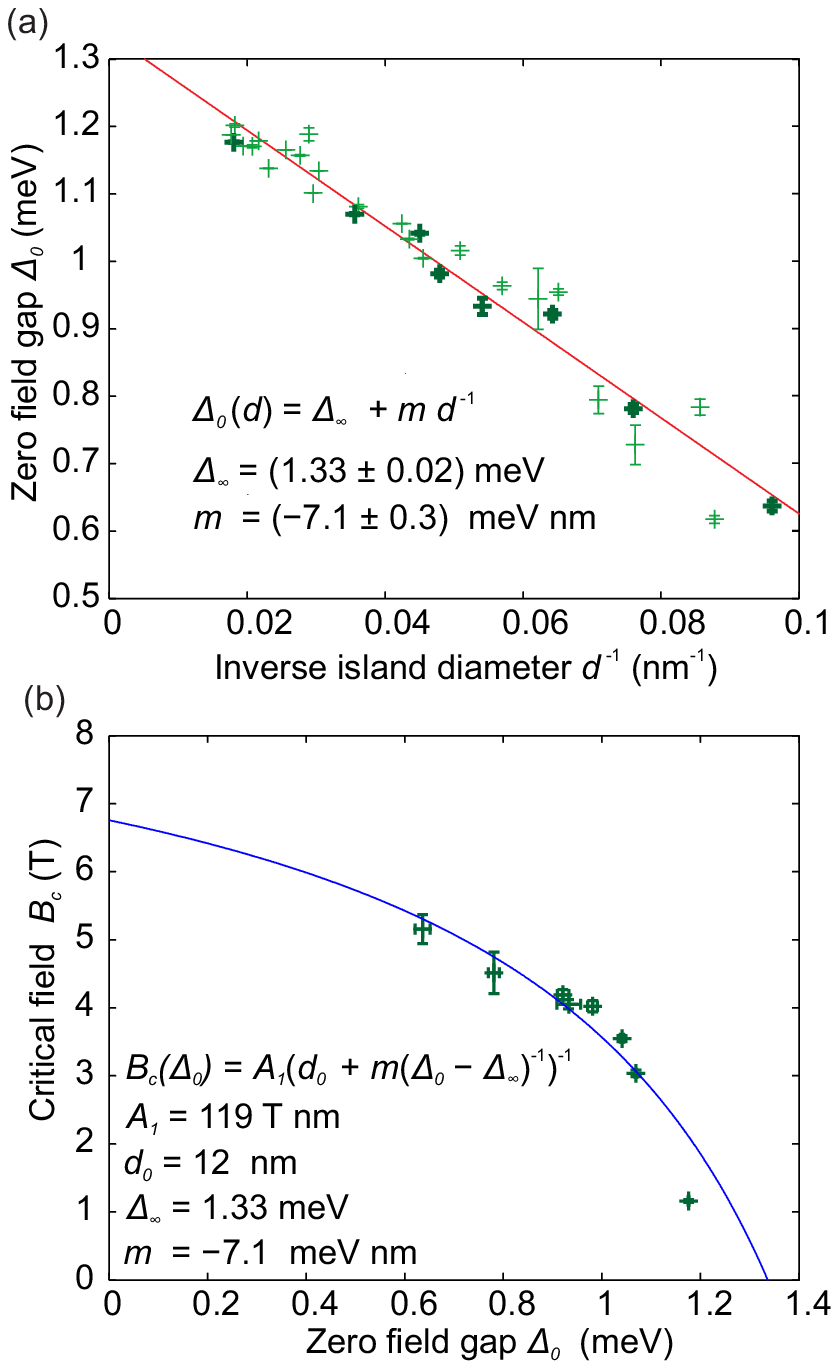}%
\caption{\label{fig4} (a) The fitted superconducting gap size $\Delta_0$ at zero magnetic field for different Pb-islands as a function of the inverse island diameter 1/d. Dark green crosses correspond to islands for which critical fields were determined in Fig. \ref{fig3}. Light green crosses note additional islands. The measurements indicate a linear dependence of the gap with the inverse island diameter $1/d$. The red line is a linear fit to $\varDelta_0 =\varDelta_\infty - m/d$. The error bars are the estimated standard deviations of the fitting parameters. (b) Critical field Bc plotted against the extracted zero-field gap $\varDelta_0$ for the islands shown in Fig. 3 (dark green crosses). This correlation of $B_c$ with $\varDelta_0$ reveals a curvature, which is also noticeable in the combination of the $A_1/(d_0+d)$ fit of the critical field $B_c(d)$ and the $m/d$ fit of the zero field gap $\varDelta_0(d)$ (blue curve). The observed behavior suggests that in the limit of vanishing superconductivity a maximal critical field of $6.7_{-1.5}^{+1.8}$  T is reached.}
\end{figure}

Despite their increased resilience to magnetic fields, smaller islands feature a reduced order parameter as signified by a reduced superconducting gap at zero field, $\varDelta_0$. We extracted $\varDelta_0$ from the $dI/dV$ spectra recorded at 0 T by fitting the Dynes formula \cite{Dynes1978,Liu2011} (see Figure S1 for details). Figure 4a shows $\varDelta_0$ as a function of island diameter which we find to fit well to $\varDelta_0 =\varDelta_\infty - m/d$. The factor $m$ sets the strength of the confinement-induced change in the zero-field gap and $\varDelta_\infty $ represents the asymptotic superconducting gap of a 7 ML thin Pb-film. A similar trend was found for the variation of $\varDelta_0$ with thickness of extended Pb islands.\cite{Brun2009} Additionally, hemispherical Pb-nanoparticles grown on  BN/Rh(111) also exhibit a similar reduction trend of the SC gap for reduced lateral dimensions.\cite{Bose2010} This indicates the generality of geometric confinement effects.  Based on the fit in Figure \ref{fig4}a we can estimate that superconductivity breaks down for islands with diameter less than $(5.3 \pm 0.2)$ nm. This result is in good agreement with theoretical predications of approximately 6.7 nm (see SI).\cite{Kim2011,Anderson1959}
\par
These measurements allow us to correlate the critical field, $B_c$, with the zero field gap, $\varDelta_0$, Figure \ref{fig4}b. Both properties are connected by their relation to the size of the Pb-island. Since decreasing island size increases the critical field and decreases the zero-field gap, the critical field is larger for islands with smaller zero-field gap. If $B_c$ and $\varDelta_0$ followed the same $1/d$ dependence the graph would show a linear trend for $B_c(0)$. But plotting $B_c$ as function of $\varDelta_0$ in Figure \ref{fig4}b shows a curvature, corroborating that critical field and zero-field gap scale differently with island size. A possible reason is proximity-induced superconductivity in the wetting layer around the Pb-islands that was discussed above. It increases the island’s effective diameter and distorts the expected $1/d$ behavior of the critical field.
\par
The critical magnetic field is limited because superconductivity vanishes for finite island sizes. The $B_c(0)$ correlation can be used to approximate the maximum achievable critical field for Pb-nanoislands. We use the fit functions $B_c(d)$ and $\varDelta_0(d)$ in Figure \ref{fig3}i and Figure \ref{fig4}a to extrapolate $B_c$ to $\varDelta_0 = 0$ meV. For the 7 ML thin Pb-islands investigated here, we estimate a maximum critical field of $6.7_{-1.5}^{+1.8}$ T for magnetic fields applied perpendicular to the surface. 
This discussion does not include the impact of the Clogston-Chandrasekar (CC) paramagnetic limit which is expected to limit the critical magnetic field to  $B_{c,limit}  =\frac{\sqrt{g}}{\mu_B} \varDelta_0$. For the smallest islands investigated in our study the limiting field would be 7.33 T which is larger than the observed critical fields. This indicates that the CC paramagnetic limit is not the dominant factor for quenching of superconductivity in Pb islands of 10 nm or more in diameter. For islands even smaller then those investigated in this study, the CC limiting field would be smaller because of the reduced gap. In such islands non-monotonic critical field behavior might be observed.
\section{CONCLUSION}
In summary we determined the critical magnetic field for nano-sized superconducting Pb-islands by directly measuring the closure of the superconducting gap as a function of magnetic field. We found critical magnetic fields as high as $(5.2 \pm 0.4)$ T for islands with 10 nm diameter. In this regime the scaling of the critical field with island diameter deviated from $1/d$ likely due to a contribution of proximity-induced superconductivity in the Pb-wetting layer. By correlating critical field and strength of superconductivity for a range of islands we estimate the maximum achievable critical field to $6.7_{-1.5}^{+1.8}$ T for islands with a diameter of $(5.3 \pm 0.2)$ nm. This highlights the possibility of tuning superconductors using spatial confinement. 

\section{METHODS}
\subsection{Sample preparation.} Pb-islands were grown on a precooled Si(111) 7$\times$7 surface at approximately 100 K to suppress diffusion and achieve small island sizes. The Si(111) 7$\times$7 surface was prepared via a series of high temperature flashes. The Si(111) surface was heated to 1170 K, and rapidly elevated several times up to 1470 K using direct-current heating. After the last flash the temperature was slowly reduced from 1220 K to room temperature with a rate of 0.5 K/sec to 1 K/sec. The high quality of the 7$\times$7 reconstruction was confirmed by STM imaging prior to Pb evaporation. Subsequently, the cold sample was placed in front of a Pb evaporator in a flux of 1.5 ML/min for 1 min. After the deposition the surface was annealed at 300 K for 10 min to allow island formation by holding the sample holder in thermal contact with the UHV chamber. After the annealing process the sample was immediately inserted back into the STM. 
\subsection{STM measurement.} The experiments were performed in an ultrahigh-vacuum low-temperature STM (type Unisoku USM-1300). All measurements were carried out at 0.6 K. The magnetic field was aligned perpendicular to the Si surface. A PtIr tip was used for the measurements. The tip was prepared by sputtering with Argon for several minutes and thermal flashing using electron-beam bombardment for several seconds. Lock-In detection was used to record the differential conductance, $dI/dV$, by adding an AC modulation voltage of 72 $\mu$eV$_{rms}$ at 730.5 Hz to the bias voltage. Tunneling spectra were recorded at a tunnel junction setpoint of 2 nA at 4 mV and the position of the tip was fixed during spectrum acquisition. All measurements were performed with a normal-conducting metal tip. We verified that the energy resolution of our STM is better than the broadening of the spectra by measuring the $dI/dV$ spectrum of a large Pb-island that showed pronounced coherence peaks (Figure S1).

\begin{acknowledgments}
The authors thank H. Suderow (Universidad Autónoma de Madrid) and Christian Ast (Max Planck Institute for Solid State Research, Stuttgart) for valuable discussions and E. Weckert and H. Dosch, (Deutsches Elektronen-Synchrotron, Germany) for providing lab space. The research leading to these results has received funding from the European Research Council (ERC-2014-StG-633818-dasQ). SRP acknowledges a fellowship from the German Academic Scholarship Foundation and JAJB postdoctoral fellowships from the Alexander von Humboldt foundation and the Natural Sciences and Engineering Research Council of Canada.
\end{acknowledgments}


\bibliography{pbislandref}

\end{document}